\documentclass[prc,aps]{revtex4}  \usepackage{graphicx} 
\begin{document} 
\draft
\title{Impact of the neutron matter equation of state on      
neutron skin and neutron drip lines in chiral effective field theory}
\author{            
Francesca Sammarruca and Yevgen Nosyk}                                                                                
\affiliation{ Physics Department, University of Idaho, Moscow, ID 83844-0903, U.S.A. 
}
\date{\today} 
\begin{abstract}
We present predictions of the binding energy per nucleon and the neutron skin
thickness in highly neutron-rich isotopes of Oxygen, Magnesium, and Aluminum. The calculations are
carried out at and below the neutron drip line as predicted 
by our model. The nuclear properties are obtained {\it via} an 
energy functional whose input is the equation of state of isospin-asymmetric infinite matter. The latter is    
based on a microscopic derivation of the energy per particle in neutron matter       
applying chiral few-nucleon forces together with a phenomenological model for the equation of state of
symmetric nuclear matter.
We highlight the impact of the neutron matter equation of state at different orders of chiral effective field theory 
on neutron skins and the binding energy per particle and quantify the uncertainty carried by our predictions. 
\end{abstract}
\maketitle

\section{Introduction} 
\label{Intro}

The behavior of the nuclear force in the medium is a complex problem. While the typical arena to test many-body 
theories is provided by finite nuclei, the system known as infinite nuclear matter is a suitable environment
to gain insight into the nature of nuclear interactions in the medium. 
Typically, nuclear matter is characterized by the energy per particle in such system, known as the equation of 
state (EoS). 
In the presence of different concentrations of neutron and protons, the symmetry energy term appears in the EoS,
whose density dependence is well known to play an outstanding role in the structure and dynamics of 
neutron-rich systems. 
This paper is part of systematic efforts in our group to explore the EoS through various applications, ranging from
neutron skins to neutron stars~\cite{Sam14}. 

Naturally, experimental constraints on the EoS are necessary, and those are extracted from measurements 
of observables which have been identified as being sensitive to the EoS (see, for instance, Ref.~\cite{Tsang+}).
Among those is 
the neutron skin thickness, which can be obtained from measurements of the neutron distribution.
Thanks to the experimental program at the Jefferson Laboratory, in the near future the weak charge density 
of some nuclei may be measured accurately.                      
In fact, 
the first of such observations was completed and yielded a value of                                                     
0.33$^{+0.16}_{-0.18}$ fm for the neutron skin thickness in 
$^{208}$Pb~\cite{Jlab}. We understand that plans are in progress to repeat the experiment aiming at a much
smaller uncertainty, and, potentially, perform a similar experiment to extract the skin of 
$^{48}$Ca~\cite{Jlab}.                                                                                       

The location of the neutron drip lines is another issue of great contemporary 
interest which is closely related to the nature of the EoS for neutron-rich matter.       
If a nucleus is extremely neutron-rich, nuclear binding may become insufficient to hold it 
together and the neutron separation energy, defined 
as $S_n = B(Z,N)-B(Z,N-1)$, where $B$ is the
binding energy, can be negative, indicating that the last neutron has become
unbound. (A similar definition applies to the proton drip lines and the
proton separation energy, but here we focus on neutron-rich systems.)
At this time, the neutron drip line is experimentally accessible only for light nuclei. However,
thanks to the recent developments of radioactive beam (RB) facilities, soon it may become possible
to explore the stability lines of nuclei ranging from light to very heavy.
Note, also, that nuclei beyond the neutron drip lines can exist in the crust of
neutron stars. Those nuclei are believed to determine, for instance, the dynamics of superfluid neutron
vortices, which, in turn, control the rotational properties of the star. 
In short, understanding the properties of nuclei with extreme neutron-to-proton ratios is an important and
challenging problem for both rare isotope beam experiments and theoretical models. 

To provide useful guidance to experiments, predictions should be                                             
accompanied by  
appropriate theoretical uncertainties. With regard to that (and more), chiral effective theory (EFT) 
\cite{Wei68,Wein79} has appealing features: it is based on the symmetries of low-energy QCD while
using degrees of freedom appropriate for low-energy nuclear physics. Furthermore, and equally important,
it allows for a systematic improvement of the predictions and a controlled
theoretical error.                                          
Therefore, in spite of the (still broad) popularity of
meson-theoretic interactions for modern calculations 
of nuclear structure and reactions,                 
chiral EFT has become established as a more fundamental and model-independent approach. In EFT,
long-range physics is determined by the interaction of pions and nucleons together with the 
(broken) symmetries of QCD, whereas 
short-range physics is included through ``contact terms" and the process of renormalization.
Together with 
an organizational scheme to rank-order the various contributions, known as power counting, 
two- and few-nucleon forces emerge on an equal footing in a controlled hierarchy.                               

In this paper, we focus on the question of how the neutron matter (NM) EoS impacts the formation of the 
neutron skin and the binding energy per particle, which, through the neutron separation
energy, determines the location of the 
drip lines, at different orders of chiral EFT.                                                                      

We will consider very neutron-rich isotopes of Oxygen, Magnesium, and Aluminum.
For the Oxygen isotopic chain, currently $^{25}$O and 
$^{26}$O are at the limit of experimental availability~\cite{oxy}, with 
$^{26}$O found to be just unbound~\cite{Kondo}.              
 With regard to Magnesium and Aluminum,
$^{40}$Mg and                                                                                                
$^{42}$Al are predicted to be drip line nuclei~\cite{Bau+,Heenen}, suggesting that the drip lines may be located 
towards heavier isotopes in this region of the nuclear chart. 

The paper is organized as follows: To ensure that the manuscript 
is self-contained, in Section~\ref{few} we give a brief review
of the few-body forces we apply, whereas in
Section~\ref{matter} we describe our approach to the energy per particle in neutron-rich matter, and 
how the latter is used in a liquid-drop-based functional in order to obtain nuclear energies and radii. 
We describe how we estimate the theoretical uncertainties in our calculations, particularly those of EFT origin.
Results and discussion are presented in Section~\ref{Res}, while our 
conclusions are summarized in Section~\ref{Concl}. 

\section{Description of the few-nucleon forces} 
\label{few}
Before moving to the description of our calculations in nucleonic matter,      
in this section we review briefly the few-nucleon forces as we will apply them in Section III.           
With regard to 3NFs, only those which do not vanish in NM will actually be employed here, although the
following remarks may be somewhat broader. 
For additional details, see Ref.~\cite{obo}.
\subsection{The chiral two-body force} 
\label{NN} 
In the present investigation we consider nucleon-nucleon (NN) potentials at order $(Q/\Lambda_\chi)^0$,
$(Q/\Lambda_\chi)^2$,
$(Q/\Lambda_\chi)^3$ and $(Q/\Lambda_\chi)^4$ in the chiral power counting, where 
$Q$ denotes the low-energy scale set by  a typical external nucleon momenta or the pion mass and 
$\Lambda_\chi$ is the chiral symmetry breaking scale. Chiral NN potentials at NLO
and N$^2$LO (order $(Q/\Lambda_\chi)^2$) and $(Q/\Lambda_\chi)^3$, respectively) 
have been constructed in Ref.~\cite{NLO} for several values of the cutoff, $\Lambda$, in the 
regulator function             
\begin{equation}
f(p',p) = exp[-(p'/\Lambda)^{2n} - 
(p/\Lambda)^{2n}] \;. 
\end{equation}
When the chiral order and the cutoff scale are changed, the low-energy 
constants in the two-nucleon sector are refitted to elastic NN scattering 
phase shifts and the deuteron properties. The low-energy constants $c_{1,3,4}$ 
associated with the $\pi \pi N N$ contact couplings of the ${\cal L}^{(2)}_{\pi N}$ chiral
Lagrangian                                 
can be extracted from $\pi N$ or NN scattering data. The potentials we use here
\cite{NLO,EM03,ME11} take the range determined 
in $\pi N$ analyses as a starting point.                                                               
The reader should consult Ref.~\cite{ME11} for details on the fitting procedure. 
    
Although 
two-body scattering phase shifts can be 
described well at NLO up to a laboratory energy of about 100 MeV~\cite{NLO} while the N$^2$LO 
potential fits the NN data up to 200 MeV, high-precision quality is not possible until             
next-to-next-to-next-to-leading order (N$^3$LO)                                            
\cite{ME11,EM03}.                    
The potential at leading order (LO)~\cite{ME11} describes NN data poorly, but we include it in
this analysis nevertheless as it may
shed additional light on the order-by-order pattern of the predictions. 

In what follows, we will employ the chiral NN potentials from Refs.~\cite{NLO,EM03,ME11} with 
cutoff parameter equal to 450 MeV and $n$=2 (for LO) or $n$=3 (for the other orders).                                             

\subsection{The chiral three-nucleon force} 
\label{3bf} 

\begin{figure}[!t] 
\centering
\includegraphics{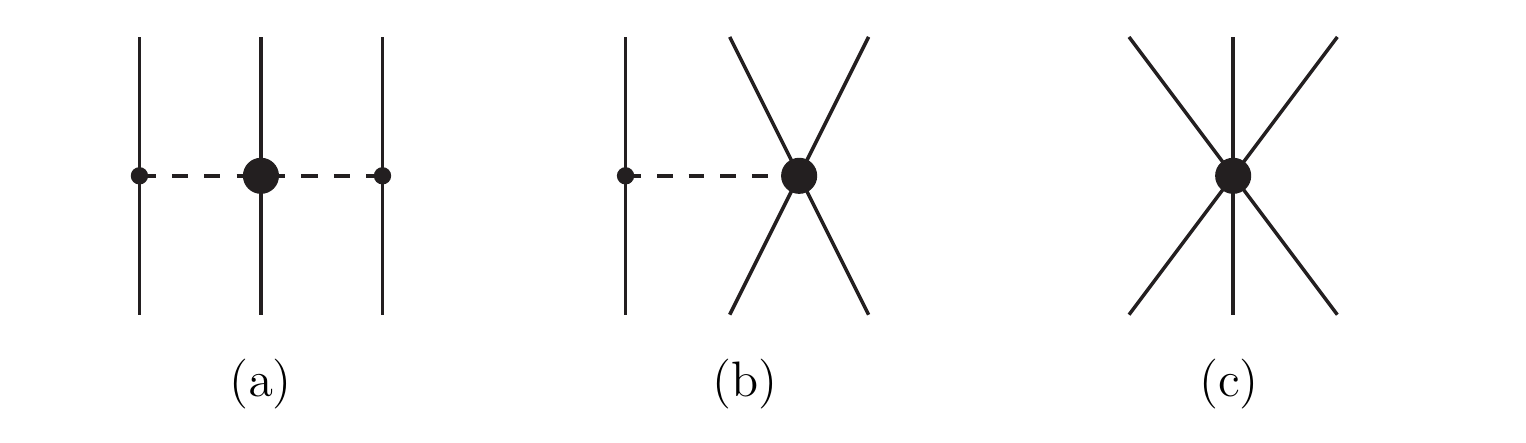}
\caption{Diagrams of the 3NF at N$^2$LO. See text for more details.} 
\label{3b}
\end{figure}

The leading three-nucleon force is encountered at third order in the chiral power 
counting and is expressed as the sum of three contributions, whose               
corresponding diagrams are shown in 
Fig.~\ref{3b}, labeled as (a), (b), (c), respectively.
These contributions are: the long-range 
two-pion-exchange part with $\pi\pi NN$ vertex proportional to the low-energy
constants $c_1,c_3,c_4$, the medium-range one-pion exchange diagram
proportional to the low-energy constant $c_D$, and the short-range contact 
term proportional to $c_E$.                                              

The inclusion of 3NFs is greatly facilitated by employing the 
density-dependent NN interaction derived in Refs.~\cite{holt09,holt10} from
the N$^2$LO chiral three-body force. This effective interaction is obtained by
summing one particle-line over the occupied states in the Fermi sea. Ignoring         
small contributions~\cite{hebeler10} depending on the center-of-mass 
momentum, the operator structure of the NN interaction is identical to the one in free 
space. 
For symmetric nuclear matter all three-body forces contribute, while for pure
neutron matter only terms proportional to the low-energy constants $c_1$ and
$c_3$ are nonvanishing~\cite{holt10,hebeler10}.                                   

While efforts are in progress to improve the status of our calculations, 
the current ``N$^3$LO'' study 
is limited to the inclusion of the N$^2$LO three-body force together with the N$^3$LO 
two-body force.                                                                     
 In Refs.~\cite{krueger,Tews}, 
calculations of the neutron matter energy per particle at N$^3$LO show a small effect 
(of about -0.5 MeV) from the N$^3$LO 3NF at saturation density for the potentials of our purview~\cite{ME11}.
Most recently, the small size of the contribution from the 3NF at N$^3$LO in NM with the potential of 
Ref.~\cite{EM03} has been confirmed~\cite{3nf}. 
The inclusion of the 3NF at N$^3$LO in nuclear matter, on the other hand, necessitates 
a refitting of the $c_D$ and $c_E$ low-energy constants, a non-trivial task still to be completed.                
With regard to the $c_i$ ($i=1,3$), for the potential under our present consideration                        
their values are $c_1=-0.81$ and $c_3=-3.40$ at both N2LO and N3LO~\cite{ME11}, as determined to best reproduce
NN data consistent with $\pi N$ analyses. The same values are used in the leading 3NF.            

It may be useful to make an additional comment concerning 
the density-dependent effective 3NF from Refs.~\cite{holt09,holt10} which we use. The latter is derived 
employing nonlocal regulators, unlike what is done when constraining the 
 $c_D$ and $c_E$ LECs from genuine 3NFs in the three-nucleon                                                      
system~\cite{Gar06,Gaz09,Marc,Viv13,Pia13,Mar13,Viv14}, a procedure which has been part of our general scheme~\cite{obo}. 
 This inconsistency, though, will 
not impact our present NM results, due to the absence of contributions proportional to 
 $c_D$ and $c_E$ in the pure neutron system.

\section{The many-body system} 
\label{matter} 
\subsection{Isospin-asymmetric matter} 
\label{NM} 

A variety of many-body methods are available and have been used extensively
to calculate the EoS of nucleonic matter.
They include: the coupled-cluster method, many-body perturbation,           
variational Monte Carlo or Green's function Monte Carlo methods. 

In computing the EoS, we employ the nonperturbative 
particle-particle (pp) ladder approximation, namely the leading
contribution in the usual hole-line expansion of the energy per particle. 
To estimate the uncertainty associated with this choice, 
in Ref.~\cite{obo} we compared              
with Refs.~\cite{cc1,cc2}                                                                           
and determined that 
the effect of using a nonperturbative approach beyond pp correlations 
is negligible in neutron matter (the focal point of
this work) and about 1 MeV per nucleon in symmetric matter around            
saturation density.

\subsection{Application in finite nuclei} 
\label{drop} 
                                                              
In order to link the EoS of asymmetric matter to an actual nucleus, we proceed as described 
in earlier work \cite{skin15}. Namely, we write the energy of a sperically symmetric nucleus {\it via} 
an energy functional based upon the semi-empirical mass formula:                         
energy of a (spherical) nucleus as 
\begin{equation}
E(Z,A) = \int d^3 r~ e(\rho,\alpha)\rho(r) + 
\int d^3 r f_0(|\nabla \rho|^2 + \beta 
|\nabla \rho_I|^2) +                 
\frac{e^2}{4 \pi \epsilon_0}(4 \pi)^2 
\int _0^{\infty} dr' r' \rho_p(r')       
\int _0^{r'} dr r^2 \rho_p(r) \; .  
\label{drop} 
\end{equation} 
Note that the integrand in the first term is 
the isospin-asymmetric equation of state,                       
\begin{equation}
e(\rho,\alpha) =                                  
e(\rho,\alpha=0) + e_{sym}\alpha^2 \; ,             
\label{eee}  
\end{equation}
with $e_{sym}$ the symmetry energy. 
In the above equation, 
$\rho$ and $\rho_I$ are defined as $\rho_n +\rho_p$ and 
$(\rho_n -\rho_p)$, respectively, and $\alpha$ represents the neutron excess,
$\alpha=\rho_I/\rho$.                                                         
We take the constant $f_0$          
in Eq.~(1) equal to 60 $MeV$ $fm^5$,                  
consistent with Ref.~\cite{Oya2010}, 
and disregard the term with the coefficient $\beta$~\cite{Furn}
because we found that its contribution was negligible.
The impact of varying the parameter $f_0$ will be addressed later.

The proton and neutron density functions are obtained by minimizing the value
of the energy, Eq.~(\ref{drop}), with respect to the paramaters of Thomas-Fermi distributions
for proton and neutron densities.  
More specifically, we write
\begin{equation}
 \rho_i(r) = \frac{\rho_0}{1 + e^{(r-a_i)/c_i}} \; , 
\label{TF}  
\end{equation}
with $i=n,p$. The radius and the diffuseness, $a_i$ and $c_i$, respectively, are optimized by minimization 
of the energy while $\rho_0$ is obtained by normalizing the proton(neutron) distribution to $Z$($N$). 
The skin is defined as 
\begin{equation}
 S = R_n - R_p \; ,                                 
\label{rnrp}   
\end{equation}
where $R_n$ and $R_p$ are the $r.m.s.$ radii of the neutron and proton density distributions, 
\begin{equation}
 R_i = \Big ( \frac{4 \pi}{T} \int_0 ^{\infty} \rho_i(r) r^4 \; dr \Big )^{1/2} \; , 
\label{rms}   
\end{equation}
where $T$= $N$ or $Z$. 
Clearly, this method is not suited to predict detailed quantum structures, such as nuclear shells or pairing effects.
On the other hand, our
purpose is not to perform detailed structure calculations, but rather to highlight the 
direct impact of the equation of state on the nuclear properties under consideration.

\begin{figure}[!t]
\vspace*{-1cm}
\includegraphics[width=7.5cm]{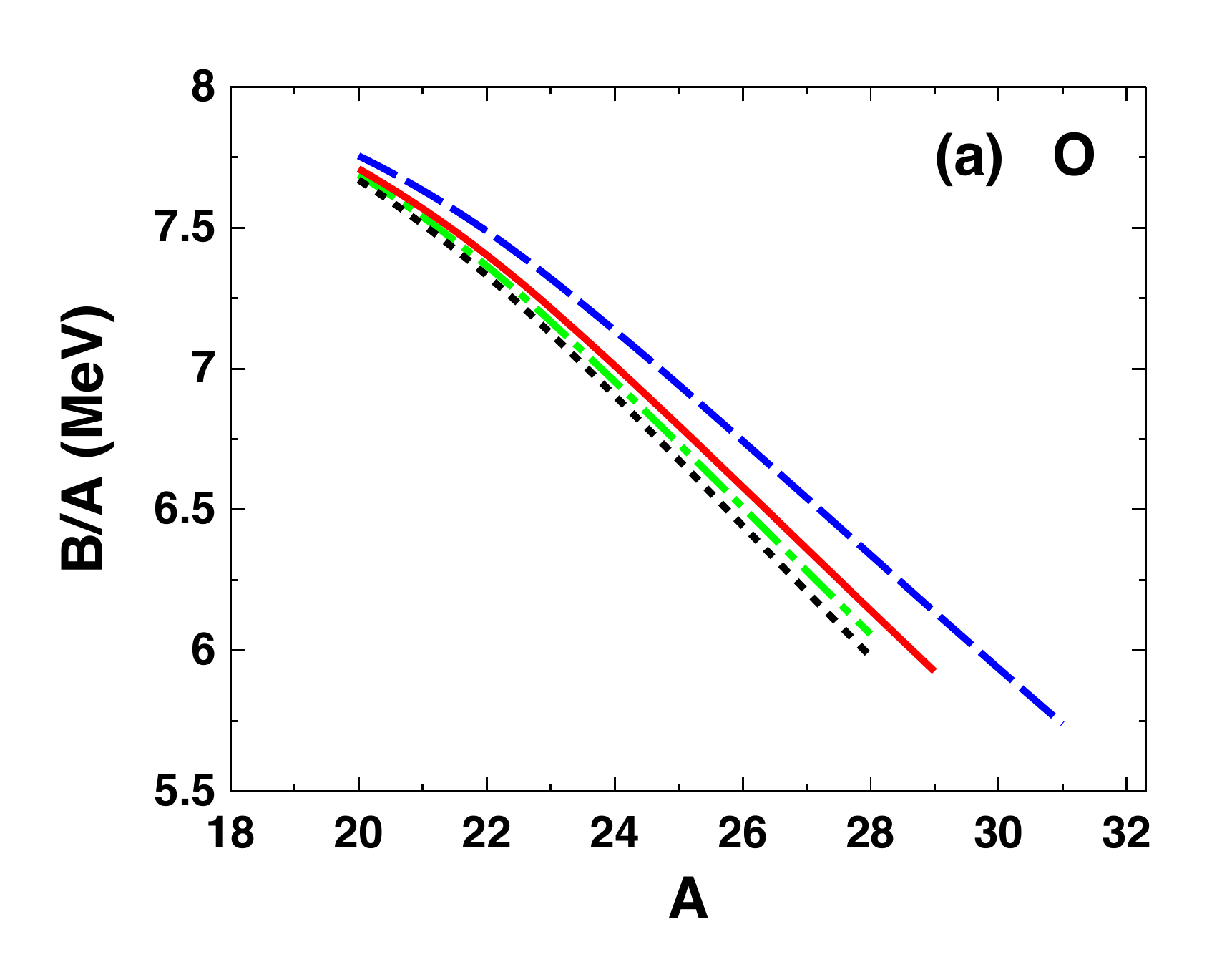}\hspace{.01in}
\includegraphics[width=7.5cm]{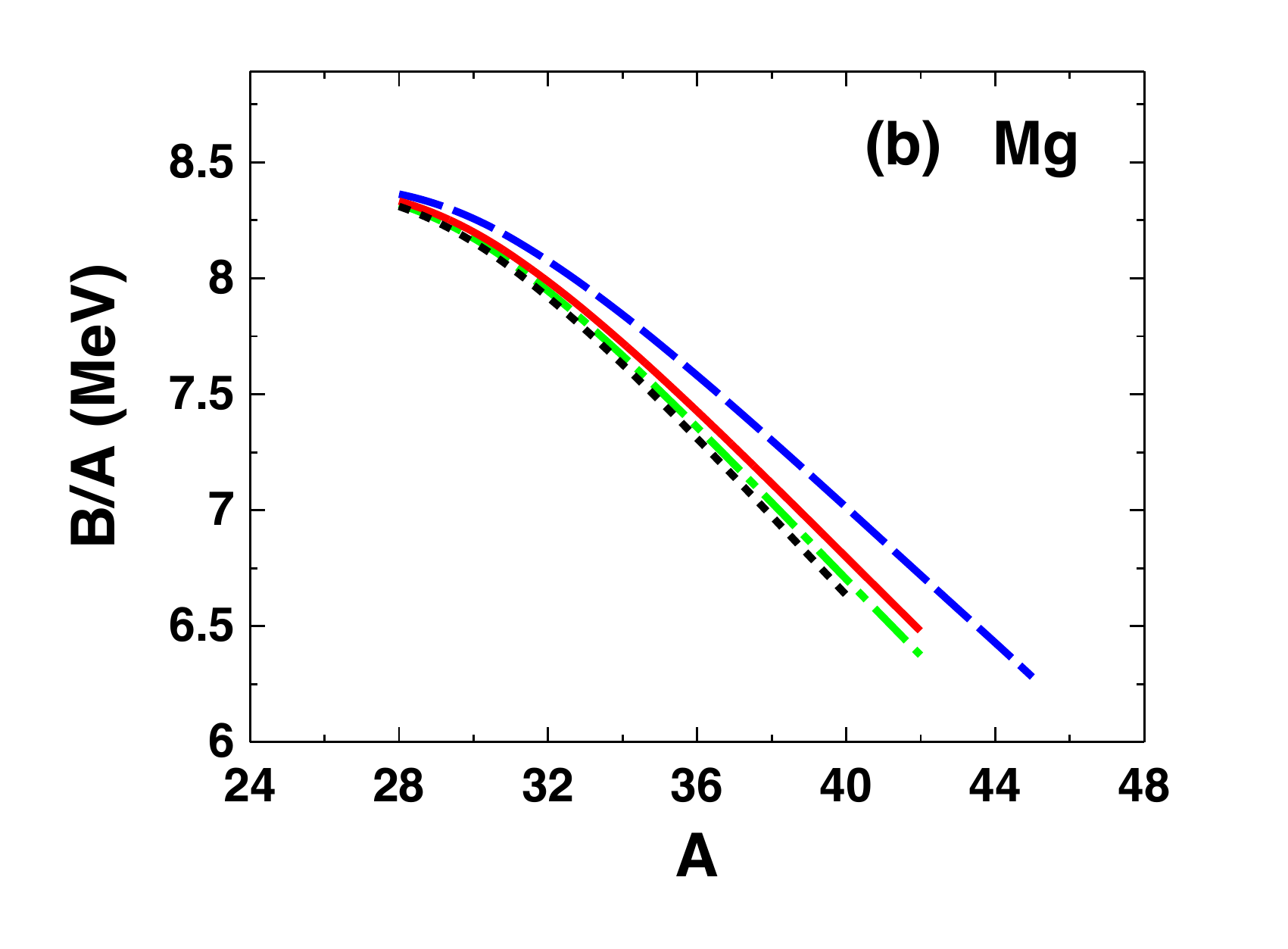}\hspace{.01in}
\includegraphics[width=7.7cm]{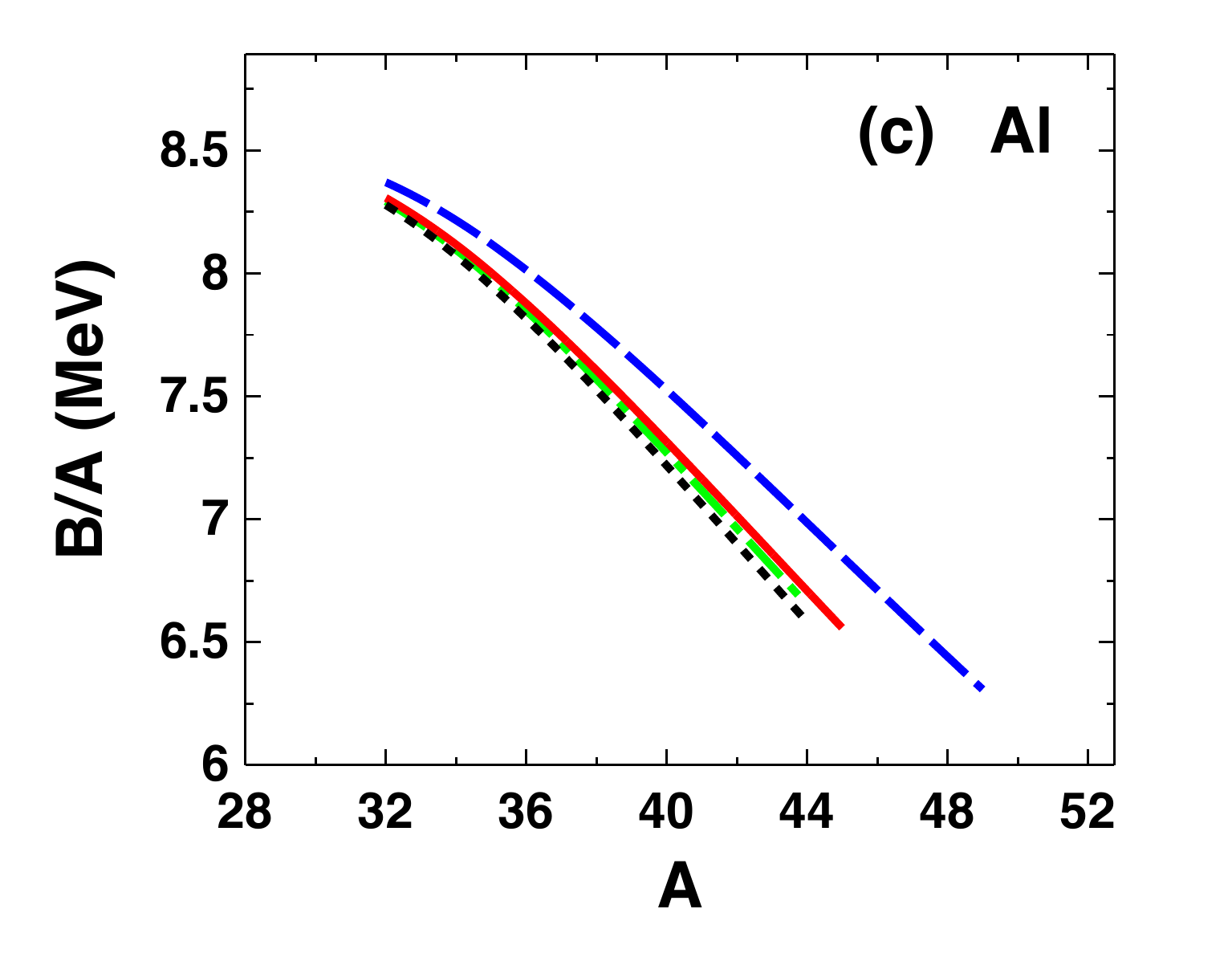}
\vspace*{0.5cm}
\caption{(Color online) (a): Binding energy per nucleon in neutron-rich isotopes 
of Oxygen {\it vs.} the mass number with increasing order of chiral EFT. Dotted black line: LO;
Dashed blue: NLO; dash-dotted green: N2LO; Solid red: N3LO. 
(b): As in (a) for Magnesium; 
(c): As in (a) for Aluminum. The various orders shown in the figure refer to the microscopic 
neutron matter equation of state, whereas a phenomenological parametrization is adopted for the 
equation of state of symmetric matter. See text for details. 
} 
\label{ba}
\end{figure}

\begin{figure}[!t]
\vspace*{1cm}
\includegraphics[width=8.5cm]{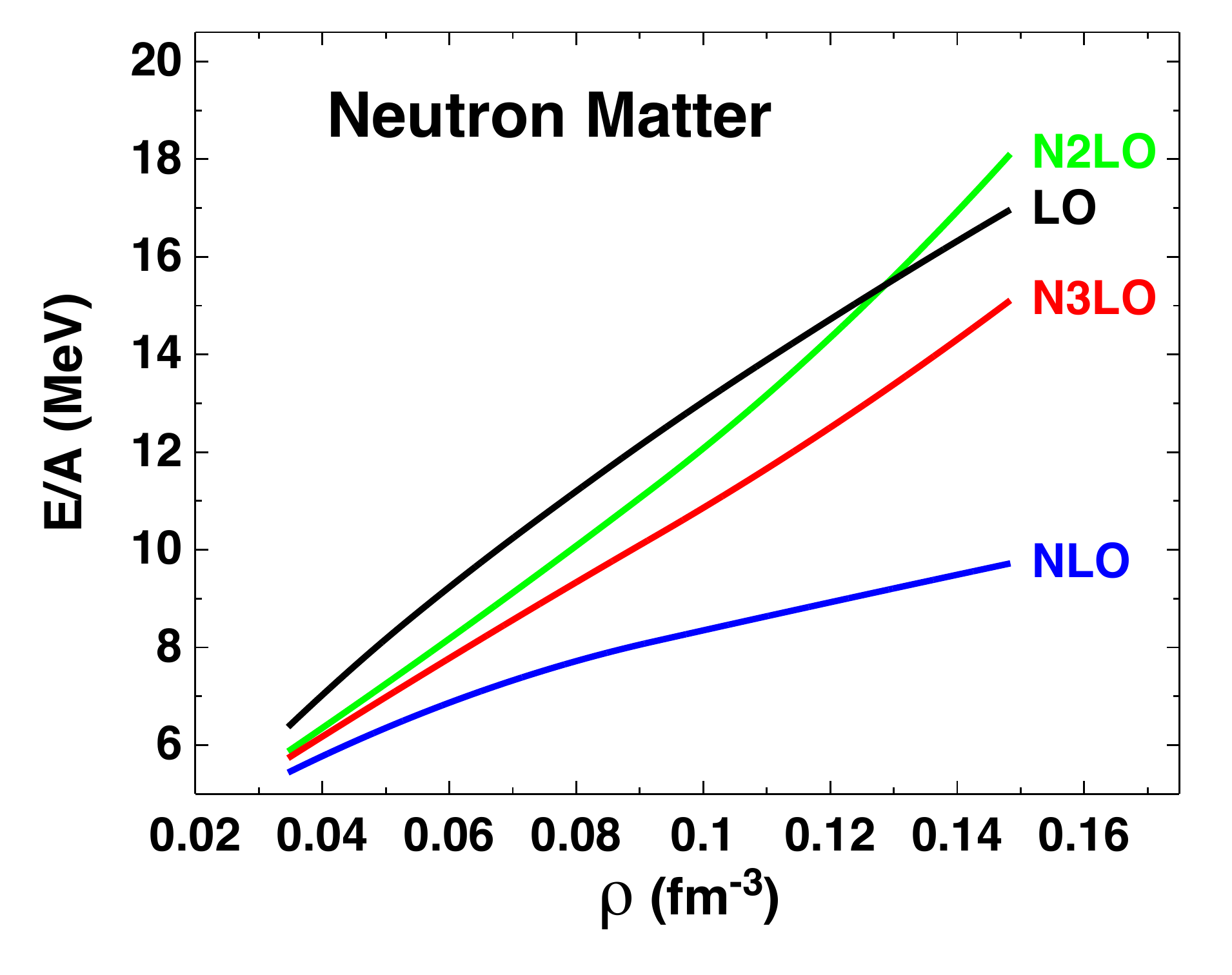}
\vspace*{0.1cm}
\caption{(Color online) The equation of state of pure neutron matter at various orders of chiral     
EFT.              
} 
\label{pnm}
\end{figure}

\begin{figure}[!t]
\vspace*{1cm}
\includegraphics[width=7.5cm]{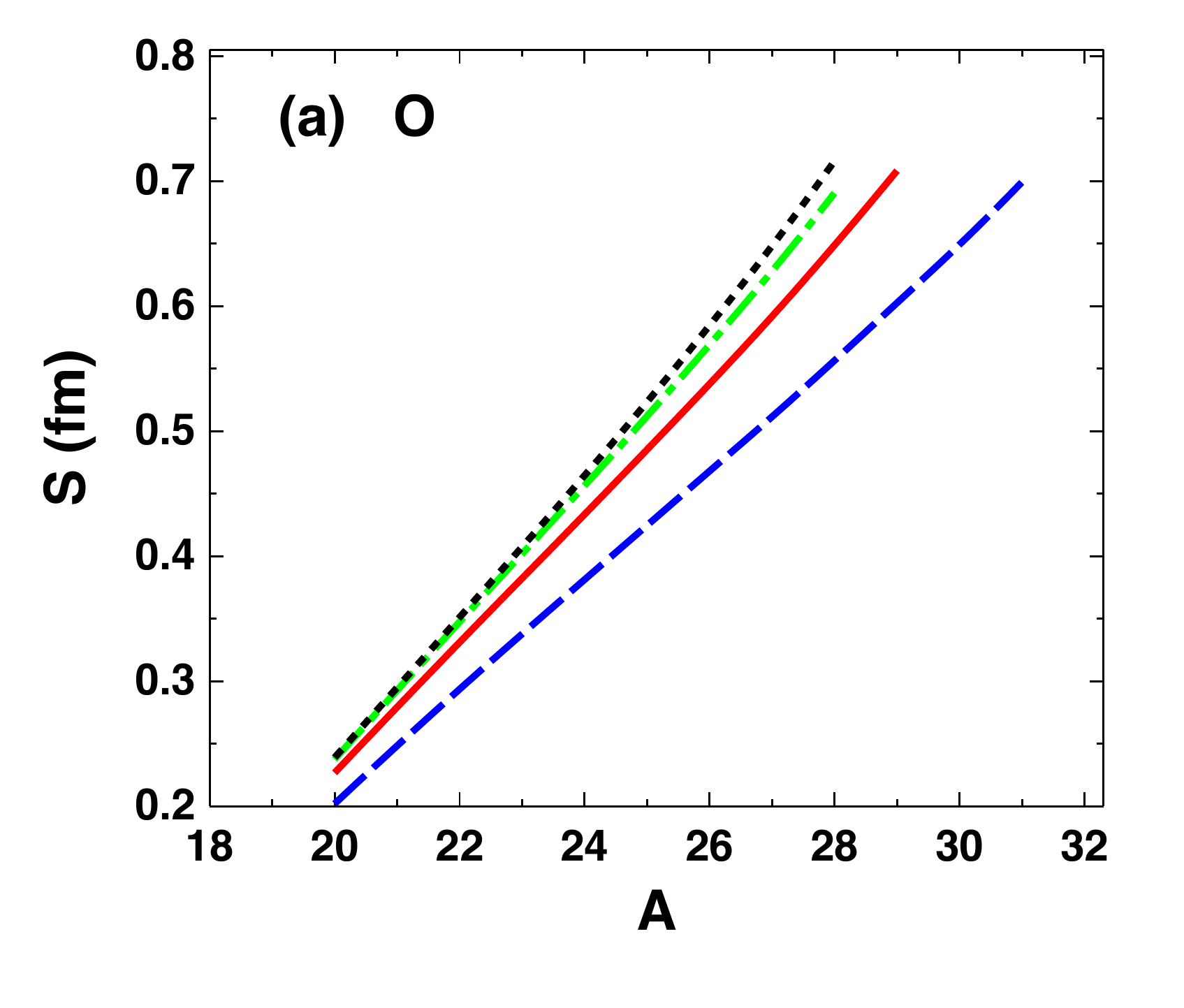}\hspace{.01in}
\includegraphics[width=7.7cm]{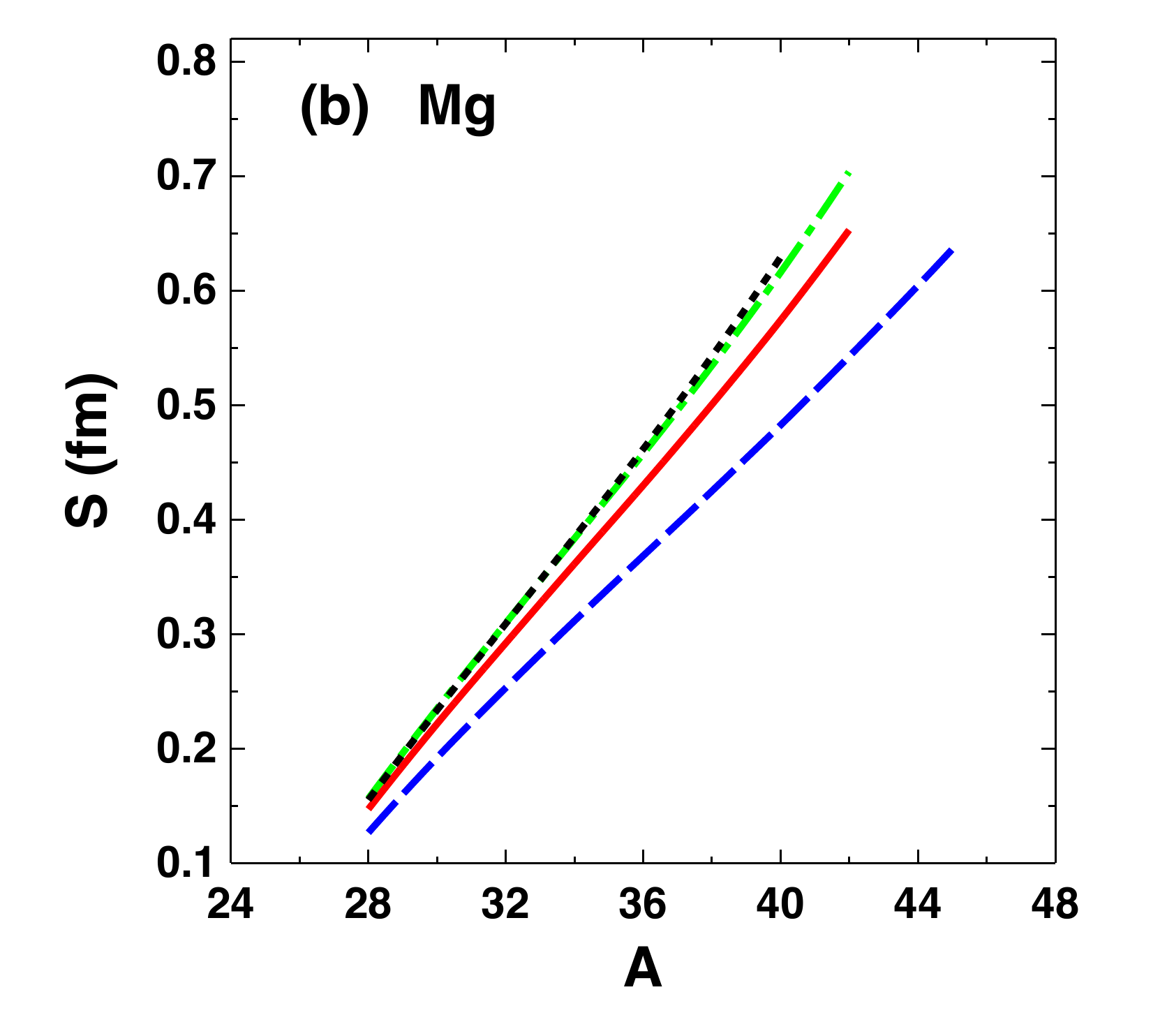}\hspace{.01in}
\includegraphics[width=7.7cm]{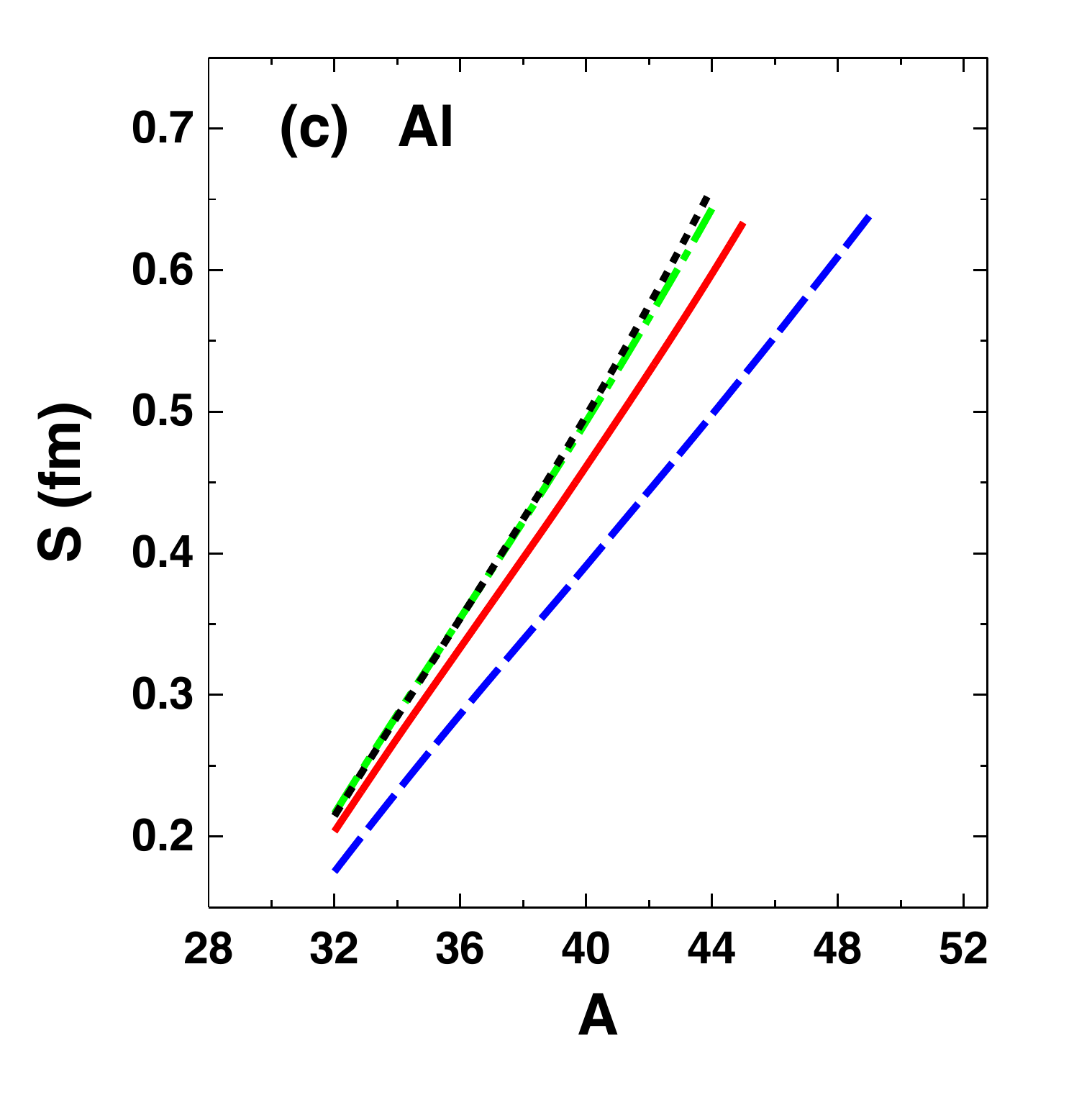}
\vspace*{0.1cm}
\caption{(Color online) As in Fig.~\ref{ba}, but for the neutron skin thickness $S$. 
} 
\label{skin}
\end{figure}

\begin{figure}[!t]
\vspace*{1cm}
\includegraphics[width=8.5cm]{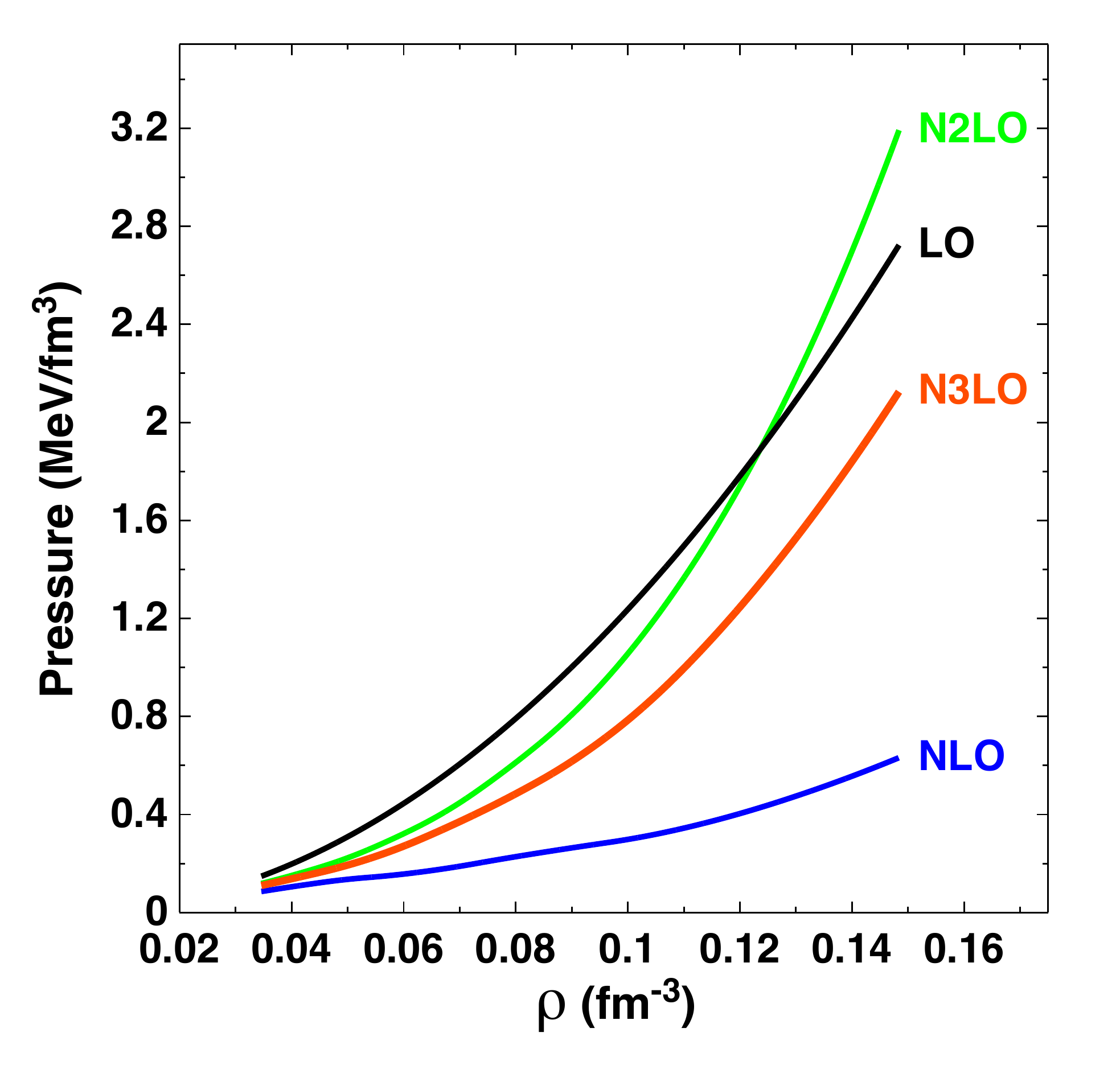}
\vspace*{0.1cm}
\caption{(Color online) The pressure in pure neutron matter for the interactions considered 
in Fig.~\ref{pnm}. 
} 
\label{prnm}
\end{figure}

\subsection{Estimation of the uncertainty} 
\label{unc} 

When addressing nucleonic matter or any other many-body system, several sources of theoretical uncertainty 
need to be considered. The one arising from the choice of the framework for obtaining the EoS was addressed 
in Sect.~\ref{NM}. Others, inherent to EFT, are: 
\begin{itemize}
\item Error in the LECs;
This item includes:          
\begin{itemize}
\item Short-range (NN) LECs;
\item Long-range ($\pi N$) LECs;
\end{itemize}
\item 
Regulator dependence;
\item Truncation error. 
\end{itemize}
We will briefly discuss each of them. 

Concerning the NN LECs, we have performed several test Brueckner-Hartree-Fock calculations 
of nucleonic matter with local high-precision
potentials from the Nijmegen group~\cite{Nij}  and concluded that the uncertainty arising from error in the   
experimental determination of NN LECs is much smaller than other errors and can be neglected. 
With regard to $\pi N$ LECs, at the 
two-body level they only impact partial waves where no contacts are available, which, at N$^3$LO,       
are $F$ waves and higher. Thus, one may expect only minor impact from this uncertainty in the two-nucleon sector.
But of course these LECs enter the 3NF, where their uncertainty can have a much larger impact. This point requires
a systematic investigation where, for each set of $c_i$ within the allowed experimental error, one constructs 
NN potentials to be used consistently in the 2NF and the 3NF. 
It is reassuring to see a recent 
Roy-Steiner analysis~\cite{Hofe} where the authors report very small errors in their determination of 
$\pi N$ LECs.                                                                                      

It has been our observation, as well as other authors'~\cite{epel+15}, that regulator dependence 
is not a good indicator of the chiral              
uncertainty at some order, as it tends to underestimate the truncation error.               
Here, we will determine the latter as explained next. 
The truncation error is essentially what is left out when terminating the chiral expansion at some order $n$.
If the prediction of observable $X$ at order $n+1$ is available, the truncation error at order $n$ is 
then 
\begin{equation}
\epsilon _n= |X_{n+1}-X_n| \; ,                                                                        
\end{equation}
which is the $(n+1)$th correction.
On the other hand, if $X_{n+1}$ is not available, the truncation error
can be estimated to be 
\begin{equation}
\epsilon _n= |X_{n}-X_{n-1}|\frac{Q}{\Lambda} \; ,      
\label{eps} 
\end{equation}
where 
$Q$ is a momentum typical for the system under consideration or the 
pion mass, and $\Lambda$ is the cutoff. Again, we have 
an expression proportional to $\Big (\frac{Q}{\Lambda} \Big )^{n+1}$.               
In our present situation, a reasonable choice for $Q$ is the Fermi momentum corresponding to the average density
of a particular nucleus. So, for each nucleus, we calculate the average density from the Thomas-Fermi 
 distributions, Eq.~(\ref{TF}), from 
which we obtain the corresponding Fermi momentum.   
The above considerations will be used in the next section to quantify the uncertainty of our predictions.

\section{Results and discussion} 
\label{Res} 

In the discussion which follows next, 
we obtain the neutron matter equation of state microscopically, as described in Ref.~\cite{obo}.
In order to emphasize the role of the pure neutron matter EoS, which is our main goal
here, we use an empirical EoS for symmetric nuclear matter (SNM) that we take from 
Ref.~\cite{snm}. 
In this way,     
we separate out the role of neutron matter pressure and remove any model dependence originating from the details of the saturation point of SNM. 

At this time we recall the remarks made at the end of Sect.~\ref{3bf} with regard to 
the contribution from the missing (N$^3$LO) 3NF expected to be very small in NM at normal density. 
Nevertheless, even with regard 
to pure neutron matter, at this stage of our calculations it is not possible to make definite statements about 
convergence of the EFT predictions from LO to N$^3$LO, since the EoS for SNM is taken from phenomenology.
When the latter, instead, is calculated microscopically, the 3NF should be obtained at 
N$^3$LO, consistent with the 2NF, in which case                                                              
the order-by-order pattern may be different than the one we see here.  
For these reasons,        
 we limit ourselves to explore the impact of the NM EoS on neutron skins and
energies at different orders while avoiding projections about convergence.                                        
Note that, for the EoS of NM, the steps from LO to N$^2$LO are free from inconsistencies. 

The phenomenological EoS of SNM                      
is obtained from a Skyrme-type energy density functional and has a realistic saturation point at $\rho_0$=0.16 fm$^{-3}$ with 
energy per particle equal to -16.0 MeV~\cite{snm}. 

Figure~\ref{ba} shows 
the binding energy per nucleon as a function of the mass number for neutron-rich isotopes of Oxygen, Magnesium, and 
Aluminum. The four curves are obtained at order 0 
(LO, dotted black), order 2 (NLO, dashed blue), order 3 (N$^2$LO,
 dash-dotted green), and order 4 (N$^3$LO, solid red) of chiral EFT in the 
calculation of the neutron matter EoS.                                       
In each case, the curves end where the neutron separation energy, $S_n = B(Z,N) - B(Z, N-1)$, turns negative.
We make the following main observations:
\begin{itemize}
\item 
The pattern shown in the figure is consistent with the change in the degree of attraction/repulsion seen in the 
NM EoS at the corresponding orders
and at the low to moderate densities
probed by the observables in this
study, see Fig.~\ref{pnm}. Namely, the most attractive interactions bind the last neutron 
up to larger values of $A$. 
\item
The order-by-order pattern is such that differences between consecutive
orders become smaller when going from LO to N$^3$LO.                  
\end{itemize}

\begin{table}                
\centering
\begin{tabular}{|c||c|c|c|}
\hline
Nucleus & Order & $B/A$ with truncation error (MeV) & $S$ with truncation error (fm) \\ 
\hline     
\hline

$^{20}$O & LO  & 7.670 $\pm$ 0.085   & 0.239 $\pm$ 0.037  \\ 
         & NLO  & 7.755 $\pm$ 0.067   & 0.202 $\pm$ 0.036  \\ 
         & N2LO  & 7.688 $\pm$ 0.021   &  0.238 $\pm$ 0.011 \\ 
         & N3LO  & 7.709 $\pm$ 0.01   &  0.227 $\pm$ 0.005 \\ 
\hline
$^{28}$O & LO  & 5.978 $\pm$ 0.361   &  0.716 $ \pm$ 0.159\\ 
         & NLO  & 6.339 $\pm$ 0.282   &  0.557 $\pm$ 0.134 \\ 
         & N2LO  & 6.057 $\pm$ 0.085   &  0.691 $\pm$ 0.042 \\ 
         & N3LO  & 6.142 $\pm$ 0.040   &  0.649 $\pm$ 0.019 \\ 
\hline
$^{28}$Mg & LO  & 8.310 $\pm$ 0.053   &  0.155 $\pm$ 0.028 \\ 
          & NLO  & 8.363 $\pm$ 0.044   &  0.127 $\pm$ 0.029 \\ 
          & N2LO  & 8.319 $\pm$ 0.014   &  0.156 $\pm$ 0.009 \\ 
          & N3LO  & 8.333 $\pm$ 0.007   &  0.147 $\pm$ 0.005 \\ 
\hline
$^{40}$Mg & LO  & 6.634 $\pm$ 0.378   &  0.621 $\pm$ 0.138 \\ 
          & NLO  & 7.012 $\pm$ 0.309   &  0.483 $\pm$ 0.133 \\ 
          & N2LO  & 6.703 $\pm$ 0.094   &  0.616 $\pm$ 0.042 \\ 
          & N3LO  & 6.797 $\pm$ 0.045   &  0.574 $\pm$ 0.020 \\ 
\hline
$^{32}$Al & LO  & 8.278 $\pm$ 0.093   &  0.215 $\pm$ 0.040 \\ 
          & NLO  & 8.371 $\pm$ 0.078   &  0.175 $\pm$ 0.041 \\ 
          & N2LO  & 8.293 $\pm$ 0.014   &   0.216 $\pm$ 0.013 \\ 
          & N3LO  & 8.307 $\pm$ 0.007   &  0.204 $\pm$ 0.007  \\ 
\hline
$^{44}$Al & LO  & 6.582 $\pm$ 0.404   &   0.659 $\pm$ 0.161 \\ 
          & NLO  & 6.986 $\pm$ 0.333   &   0.498 $\pm$ 0.146 \\ 
          & N2LO  & 6.653 $\pm$ 0.082   &   0.644 $\pm$ 0.046\\ 
          & N3LO  & 6.709 $\pm$ 0.027   &   0.598 $\pm$ 0.022 \\ 
\hline
\end{tabular}
\caption                                                    
{ Binding energy per nucleon ($B/A$)                                                  
and neutron skin ($S$), along 
with their truncation error at each order, for some of the neutron-rich nuclei from 
Figs.~\ref{ba},\ref{skin}. See text for more details.                                     
} 
\label{tab1}
\end{table}

\begin{table}                
\centering
\begin{tabular}{|c|c|c|c|}
\hline
$ k_F^n$ (fm$^{-1}$) & Order & $E/A$ with truncation error (MeV)  \\ 
\hline     
1.39  & LO & 12.126 $\pm$ 4.10   \\ 
      & NLO & 8.027 $\pm$ 2.99   \\ 
    & N2LO & 11.017 $\pm$ 0.95   \\ 
    & N3LO & 10.063 $\pm$ 0.58   \\ 
\hline     
\end{tabular}
\caption                                                    
{The energy per neutron in NM ($E/A$)                                                  
with its truncation error at the 
indicated chiral orders. The value of the {\it neutron} Fermi momentum, $k_F^n$, corresponds approximately to 
the average density determined earlier for the nuclei under consideration.                                   
} 
\label{tab2}
\end{table}

\begin{table}                
\centering
\begin{tabular}{|c||c|c|c|}
\hline
Nucleus & Order & $B/A$ (MeV) & $S$ (fm) \\ 
\hline     
\hline
$^{20}$O & LO  & 7.445 $\pm$ 0.226   & 0.248 $\pm$ 0.009  \\ 
         & NLO  & 7.526 $\pm$ 0.230   & 0.211 $\pm$ 0.009  \\ 
         & N2LO  & 7.463 $\pm$ 0.225   &  0.246 $\pm$ 0.008 \\ 
         & N3LO  & 7.483 $\pm$ 0.227  &  0.236 $\pm$ 0.009 \\ 
\hline
$^{28}$O & LO  & 5.825 $\pm$ 0.153   &  0.740 $\pm$ 0.024 \\ 
         & NLO  & 6.170 $\pm$ 0.170   &  0.581 $\pm$ 0.024 \\ 
         & N2LO  & 5.904 $\pm$ 0.153   &  0.712 $\pm$ 0.021 \\ 
         & N3LO  & 5.985 $\pm$ 0.158   &  0.671 $\pm$ 0.022 \\ 
\hline
$^{28}$Mg & LO  & 8.094 $\pm$ 0.216   &  0.162 $\pm$ 0.006 \\ 
          & NLO  & 8.145 $\pm$ 0.218   &  0.133 $\pm$ 0.006 \\ 
          & N2LO  & 8.110 $\pm$ 0.210   &  0.162 $\pm$ 0.006 \\ 
          & N3LO  & 8.117  $\pm$ 0.216   &  0.153 $\pm$ 0.006 \\ 
\hline
$^{40}$Mg & LO  & 6.489 $\pm$ 0.146   &  0.652 $\pm$ 0.022 \\ 
          & NLO  & 6.851 $\pm$ 0.161   &  0.504 $\pm$ 0.022 \\ 
          & N2LO  & 6.558 $\pm$ 0.145   &  0.636 $\pm$ 0.020 \\ 
          & N3LO  & 6.647 $\pm$ 0.150   &  0.595 $\pm$ 0.021 \\ 
\hline
$^{32}$Al & LO  & 8.075 $\pm$ 0.203   &  0.223$\pm$ 0.009 \\ 
          & NLO  & 8.164 $\pm$ 0.207   &  0.183 $\pm$ 0.008 \\ 
          & N2LO  & 8.091 $\pm$ 0.203   &   0.224 $\pm$ 0.008 \\ 
          & N3LO  & 8.108 $\pm$ 0.199   &  0.212 $\pm$ 0.008  \\ 
\hline
$^{44}$Al & LO  & 6.443 $\pm$ 0.139   &   0.682 $\pm$ 0.023 \\ 
          & NLO  & 6.831 $\pm$ 0.155   &   0.521 $\pm$ 0.023 \\ 
          & N2LO  & 6.515 $\pm$ 0.138   &   0.665 $\pm$ 0.021\\ 
          & N3LO  & 6.589 $\pm$ 0.121   &   0.620 $\pm$ 0.022 \\ 
\hline
\end{tabular}
\caption                                                    
{ Binding energy per nucleon ($B/A$)                                                  
and neutron skin ($S$) for the same nuclei considered in Table~\ref{tab1}. 
The values shown are an average of the predictions obtained with $f_0$= 60 MeV fm$^5$ and 
those obtained with $f_0$= 70 MeV fm$^5$~\cite{Oya2010} with the error arising from such variation. 
} 
\label{tab3}
\end{table}

\begin{table}                
\centering
\begin{tabular}{|c||c|c|c|}
\hline
Nucleus & Order & $B/A$ (MeV) & $S$ (fm) \\ 
\hline     
\hline
$^{20}$O & LO  &                                              
 7.445 $\pm$ 0.241 & 0.248 $\pm$ 0.038 \\
         & NLO  &                                              
 7.526 $\pm$ 0.240 & 0.211 $\pm$ 0.037 \\
 & N2LO  &                                              
 7.463 $\pm$ 0.226 & 0.246 $\pm$ 0.014 \\
 & N3LO  &                                              
 7.483 $\pm$ 0.227 & 0.236 $\pm$ 0.010\\
\hline
$^{28}$O & LO  &                                              
 5.825 $\pm$ 0.392 & 0.740 $\pm$ 0.161 \\
         & NLO  &                                              
 6.170 $\pm$ 0.329 & 0.581 $\pm$ 0.136 \\
 & N2LO  &                                              
 5.904 $\pm$ 0.175 & 0.712 $\pm$ 0.047 \\
 & N3LO  &                                              
 5.985 $\pm$ 0.163 & 0.671 $\pm$ 0.029 \\
\hline
$^{28}$Mg & LO  &                                              
 8.094 $\pm$ 0.222 & 0.162 $\pm$ 0.029 \\
          & NLO  &                                              
 8.145 $\pm$ 0.222 & 0.133 $\pm$ 0.031 \\
 & N2LO  &                                              
 8.110 $\pm$ 0.210 & 0.162 $\pm$ 0.011 \\
 & N3LO  &                                              
 8.117 $\pm$ 0.216 & 0.153 $\pm$ 0.008 \\
\hline
$^{40}$Mg & LO  &                                              
 6.489 $\pm$ 0.405 & 0.652 $\pm$ 0.140 \\
          & NLO  &                                              
 6.851 $\pm$ 0.348 & 0.504 $\pm$ 0.135 \\
 & N2LO  &                                              
 6.558 $\pm$ 0.173 & 0.636 $\pm$ 0.047 \\
 & N3LO  &                                              
 6.647 $\pm$ 0.157 & 0.595 $\pm$ 0.029 \\
\hline
$^{32}$Al & LO  &                                              
 8.075 $\pm$ 0.223 & 0.223 $\pm$ 0.041 \\
          & NLO  &                                              
 8.164 $\pm$ 0.221 & 0.183 $\pm$ 0.042 \\
 & N2LO  &                                              
 8.091 $\pm$ 0.203 & 0.224 $\pm$ 0.015 \\
 & N3LO  &                                              
 8.108 $\pm$ 0.199 & 0.212 $\pm$ 0.011 \\
\hline
$^{44}$Al & LO  &                                              
 6.443 $\pm$ 0.427 & 0.682 $\pm$ 0.163 \\
          & NLO  &                                              
 6.831 $\pm$ 0.367 & 0.521 $\pm$ 0.148 \\
 & N2LO  &                                              
 6.515 $\pm$ 0.161 & 0.665 $\pm$ 0.051 \\
 & N3LO  &                                              
 6.589 $\pm$ 0.124 & 0.620 $\pm$ 0.031 \\
\hline
\end{tabular}
\caption                                                    
{ Binding energy per nucleon ($B/A$)                                                  
and neutron skin ($S$) for the same nuclei as in Table~\ref{tab1} with their compounded error. 
} 
\label{tab4}
\end{table}

In Fig.~\ref{skin}, the neutron skin thickness is shown as a function of $A$ for the same isotopes
and physical interactions as considered in Fig.~\ref{ba}. The 
largest values of the skin are obtained with the most repulsive NM EoS. Note also how the pressure in 
neutron matter at the various orders shown in Fig.~\ref{prnm} reveals large differences among the 
various interactions.
The order-by-order pattern is consistent with what we observed for the binding energy.
At the low densities (likely to be probed by the skin), the LO and the N$^2$LO interactions yield very 
similar values for the pressure and the skin.

We provide additional information in Table~\ref{tab1}, where we made some selections in order 
to avoid an excessively cumbersome tabulation. For each of the three elements under consideration, we show
the smallest value of $A$ from Figs.~\ref{ba},\ref{skin} and the value of $A$ (also 
from Figs.~\ref{ba},\ref{skin})                                     
for which the separation energy first becomes negative, namely, the first 
value of $A$ at which one of the three curves is interrupted. (This way, the energies 
and skins in the tables are comparable with one another order by order, whereas the ``drip" $A$ would be different
at each order.) 
The emerging pattern is clear
and suggests that the truncation error decreases at the higher orders of the
expansion, for both the energy and the skin. Also, at fixed order, the uncertainty is larger for the more
neutron-rich systems, most likely 
reflecting  increasing role of the microscopic 
NM EoS with its corresponding uncertainty.                               

Concerning the latter, we provide some quantitative information in Table~\ref{tab2}. 
As described at the end of Sec.~\ref{unc}, 
in developing Table~\ref{tab1} we needed to find an average density relevant 
for the nuclei included in this investigation. The latter was found
to range from about 0.081 to 0.095 fm$^{-3}$, or, in terms of the Fermi momentum of SNM,
from about 1.06 to 1.12 fm$^{-1}$. Thus,                                         
$k_F$ =1.1 fm$^{-1}$ is representative, which translates into       
the {\it neutron} Fermi momentum entered in Table~\ref{tab2}. The Table shows the energy per neutron at the various orders
with their truncation error, which has been calculated as explained earlier, with $k_F^n$ as the typical 
momentum used in Eq.~(\ref{eps}). 

Finally we like to discuss an uncertainty that is not EFT-related. This is due 
to variations of the $f_0$ 
parameter~\cite{Oya2010} in the surface term of the liquid droplet functional, Eq.~(\ref{drop}), when fitted to $\beta$-stable nuclei. This uncertainty is 
displayed in Table~\ref{tab3}.                         
The values shown are an average of the predictions obtained with $f_0$= 60 MeV fm$^5$ and 
those with $f_0$= 70 MeV fm$^5$~\cite{Oya2010} with the error arising from such variation. 
We see that the uncertainty associated with this parameter                                     
is approximately independent of the chiral order. For the energy, 
it is generally larger than the truncation error, 
although the latter may dominate at LO. For the skin, on the 
other hand, it is smaller than or comparable with the truncation error, particularly at the highest order. 
Note that the error displayed in Table~\ref{tab3} is not related to the Hamiltonian, whose
pattern by chiral order remains the same regardless the value of $f_0$. 
The final results including their compounded error (calculated in quadrature) are shown in Table~\ref{tab4}.

It is also interesting to
mention recent {\it ab initio} calculations of medium-mass neutron-rich 
nuclei, $^{48}$Ca in particular~\cite{hagen+}. There, the neutron skin thickness 
in $^{48}$Ca was predicted with various low-momentum chiral Hamiltonians~\cite{Heb11} and found to be nearly independent
of the interaction, due to a strong correlation between the point neutron and proton
radii.          
Here, we have considered a group of interactions at different chiral                    
orders while keeping the properties of SNM fixed.          
Under the present circumstances, we find that  
larger NM pressure corresponds to larger neutron skin.                                                                 

\begin{figure}[!t]
\vspace*{1cm}
\includegraphics[width=6.5cm]{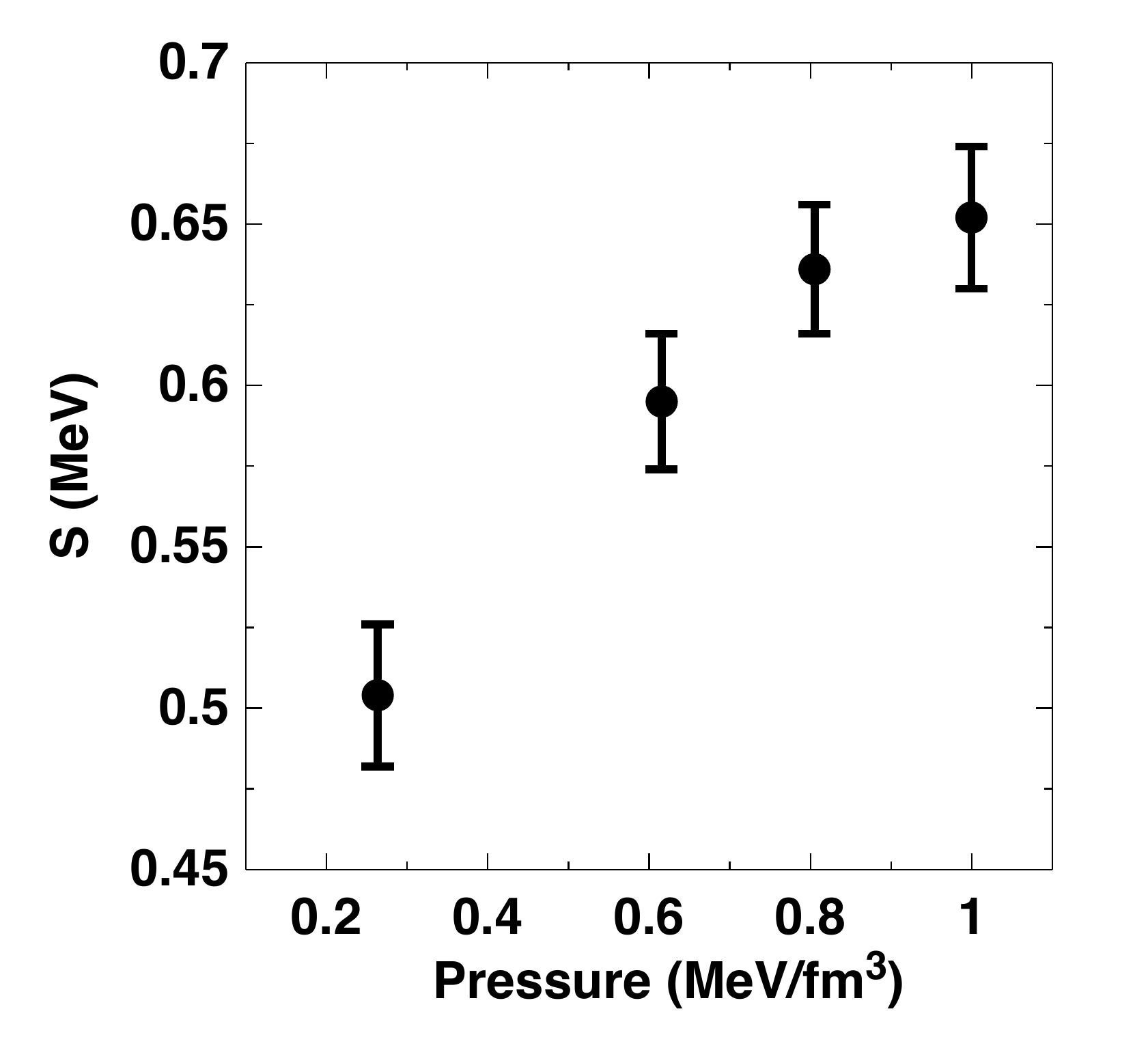}
\includegraphics[width=6.5cm]{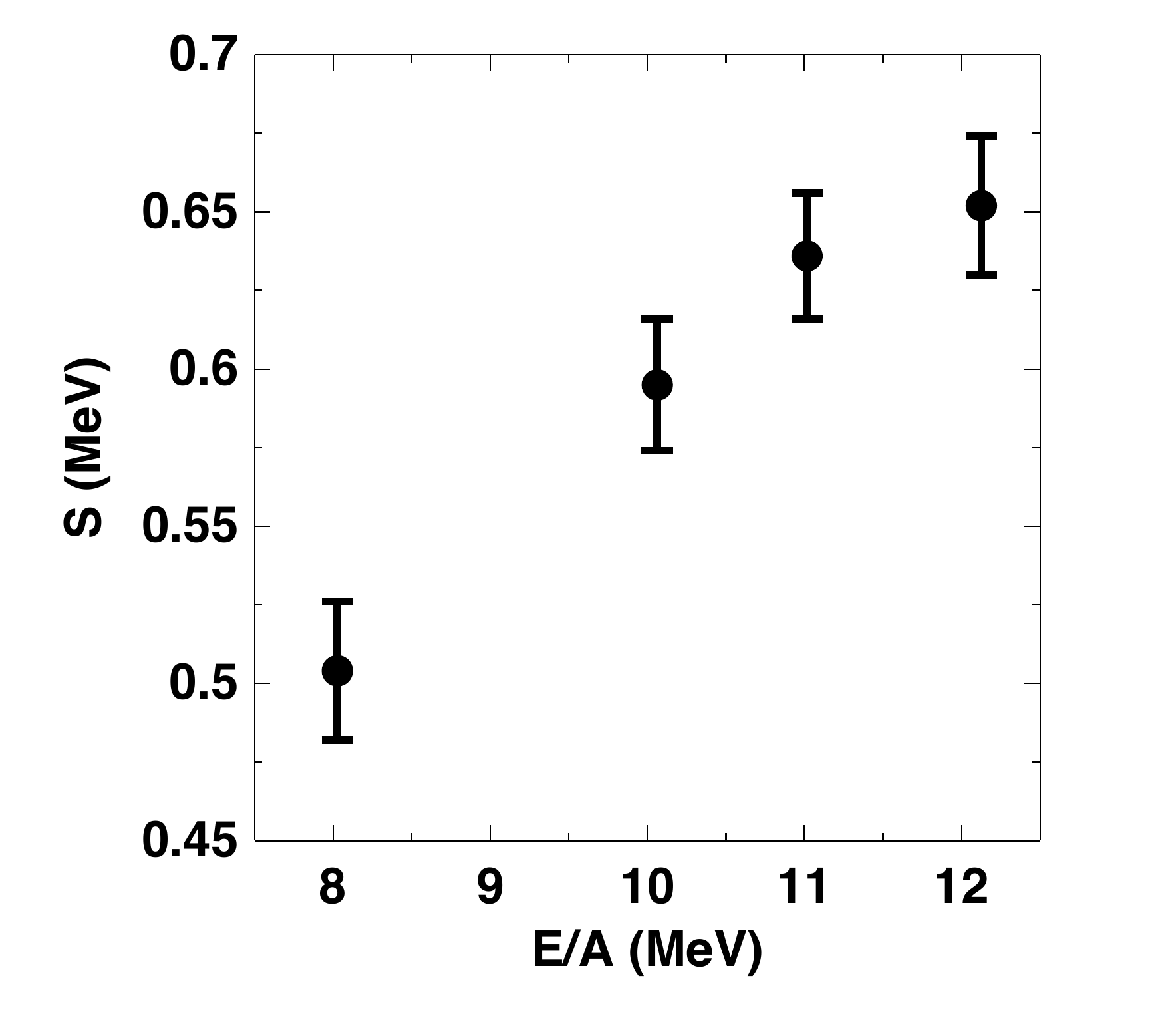}
\vspace*{0.1cm}
\caption{(Color online) Correlation between the skin, $S$, and the pressure (left) or   
the energy in neutron matter (right). The density is fixed and equal to the average
density in 
$^{40}$Mg. Skin values as in Table~\ref{tab3}. 
} 
\label{pes}
\end{figure}

Before closing, we wish to extend the discussion and explore                        
correlations among the main quantities addressed in this investigation. Linear correlations between
two variables are usually studied 
using the Pearson coefficient:                                          
\begin{equation}
\rho(x,y) = \frac{cov(x,y)}{\sigma_x \sigma_y} \;, 
\end{equation}
where the covariance $cov(x,y)$ is defined as
\begin{equation}
cov(x,y) = \sum_{i=1}^{n}\frac{(x_i-{\bar x})(y_i-{\bar y})}{ n-1} \;,
\end{equation}
and ${\bar x}$ and ${\bar y}$ are the average values of the $\{x_i\}$ and $\{y_i\}$ data sets, respectively.
${\sigma_x}$ and ${\sigma_y}$ are the usual standard deviations. 
Note that our samples contain only four data, namely the skins at 
LO to N$^3$LO and the pressure or energy at the corresponding orders (at some chosen, fixed density). This  
may render the Pearson coefficient, or the identification of a specific fitting
function, unreliable.                  
We will show correlations graphically, see Fig.~\ref{pes}. There, we see a positive correlation between skin and either the pressure or the energy per particle in NM.
The figure shows the skin of $^{40}$Mg, but the behavior is representative for the other nuclei. 

\begin{figure}[!t]
\vspace*{1cm}
\includegraphics[width=6.5cm]{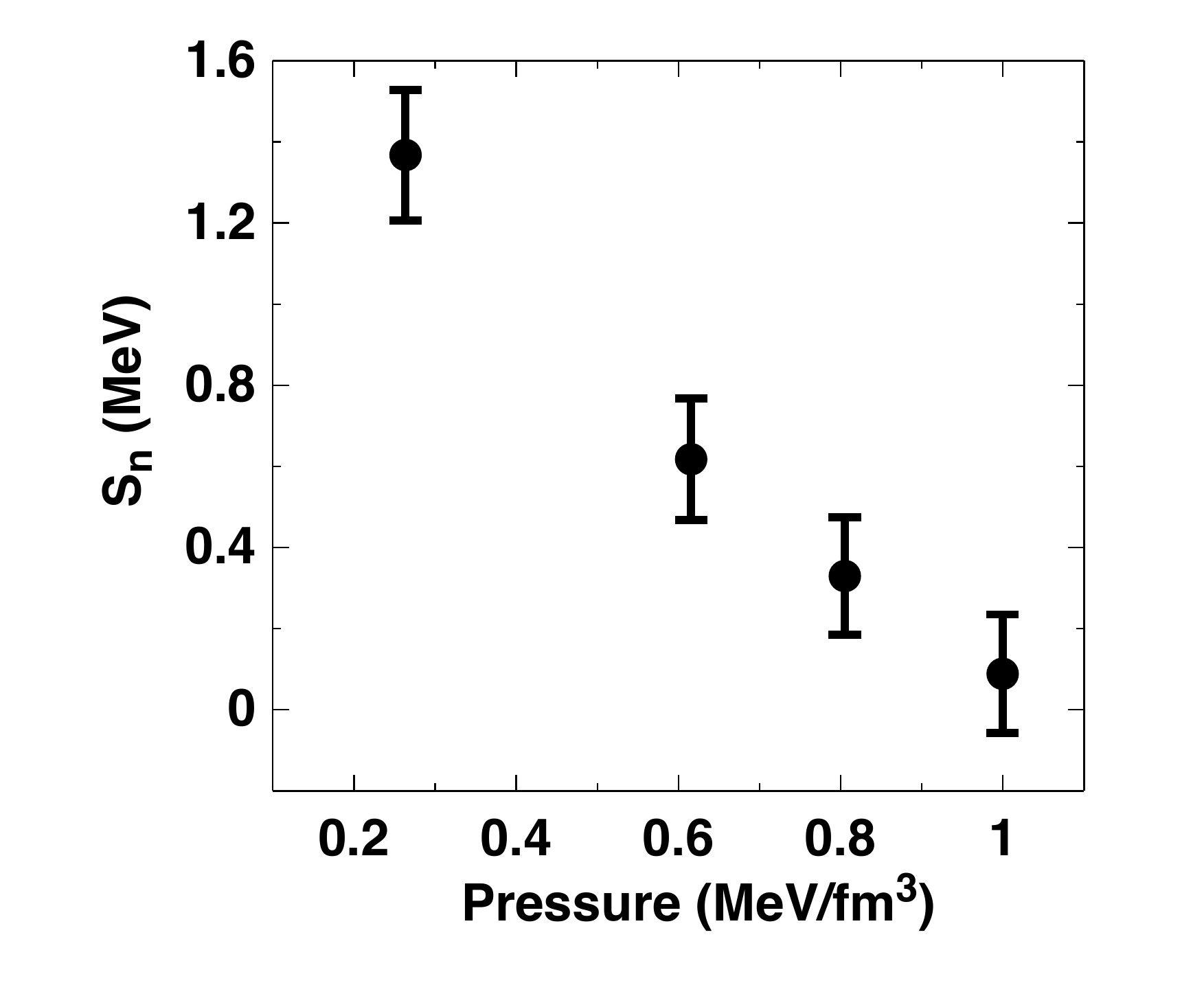}
\includegraphics[width=5.8cm]{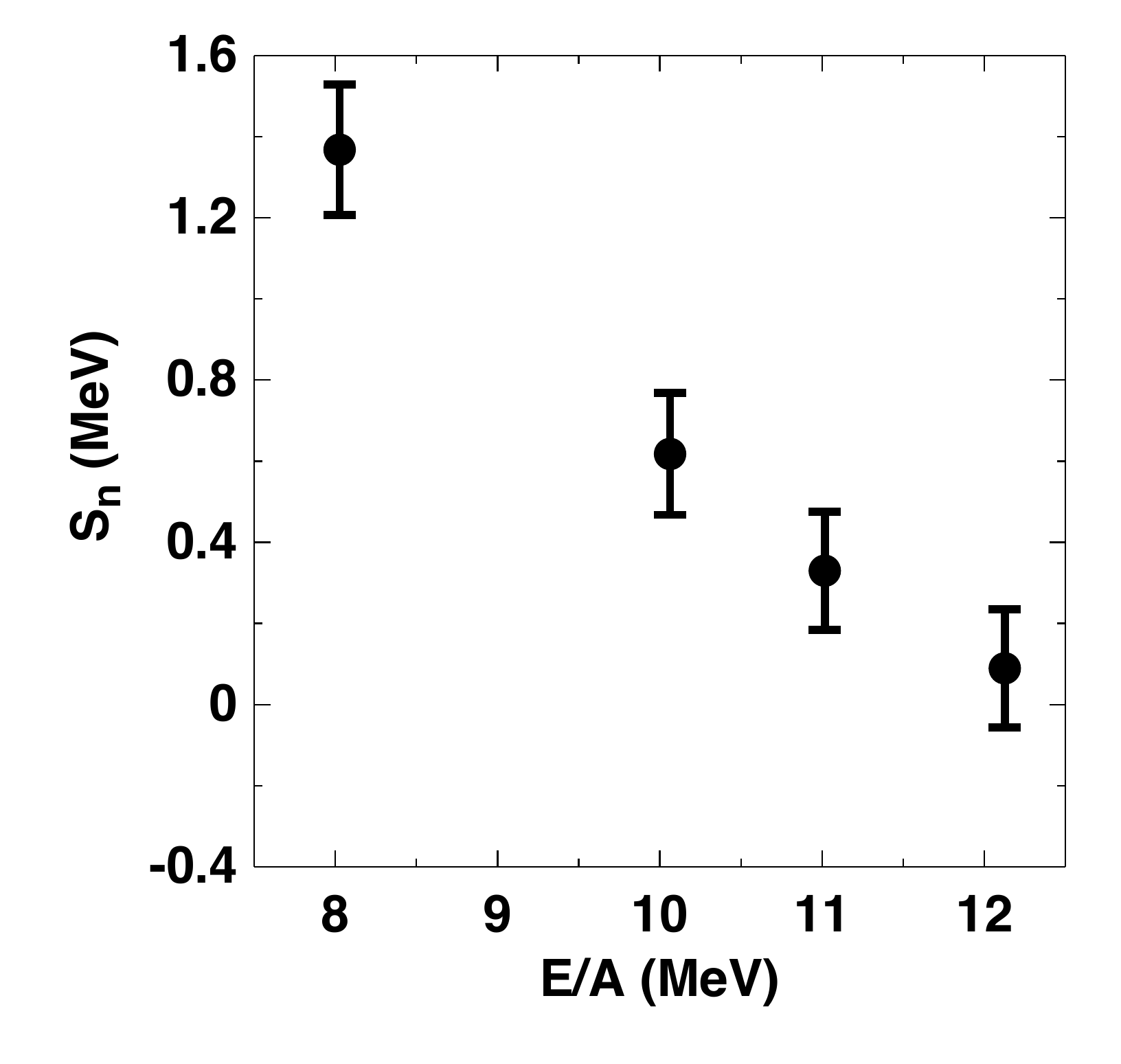}
\vspace*{0.1cm}
\caption{(Color online) Correlation between the neutron separation 
energy, $S_n$, and the pressure (left) or  
the energy in neutron matter (right). Details are given in the text.
} 
\label{pesep}
\end{figure}

To make a similar analysis of how the drip lines correlate with neutron matter pressure or energy, 
we consider the neutron separation energy,
 $S_n = B(Z,N)-B(Z,N-1)$, which is chiefly responsible for the location of the driplines.
For a particular nucleus,                   
again $^{40}$Mg, we calculate          
 $S_n = B(Z=12,N=28)-B(Z=12,N=27)$.                                                                  
The correlation between 
the separation energy at the four orders and the corresponding pressure and energy in NM are shown in 
in Fig.~\ref{pesep} on the left side and the right side, respectively. 
We see a definite anticorrelation of the neutron removal energy with either pressure or energy in NM. 
Since a smaller separation energy signifies  that the drip line is closer, we conclude that 
either larger NM pressure or larger energy will facilitate the onset of the drip lines. 

We like to end with a comparison with currently available empirical information.
As mentioned in the Introduction, 
experimental information on very neutron-rich nuclei, particularly neutron densities,
is still scarse, a state of affairs which is expected to improve with measurements at RB facilities and 
the electroweak program at Jlab. 
To gain a better insight on how the predictions from our functional compare with available tabulations, experimental or
estimated, we consulted the large compilation of nuclear data from Ref.~\cite{Ame2012}. The values we found for the 
binding energy per nucleon compare favorably with those in Table~\ref{tab4}. For instance, Ref.~\cite{Ame2012} reports
for $^{28}$Mg a value of 8.2724 MeV (our predictions at N$^3$LO: 8.117 $\pm$ 0.216 MeV); 
for $^{40}$Mg a value of 6.621 MeV (our predictions at N$^3$LO: 6.647 $\pm$ 0.157 MeV); 
for $^{20}$O a value of 7.568 MeV (our predictions at N$^3$LO: 7.483 $\pm$ 0.227 MeV); 
for $^{32}$Al a value of 8.100 MeV (our predictions at N$^3$LO: 8.108 $\pm$ 0.199 MeV). 

As another test of the general validity of the functional method (regardless the EoS), we calculated 
the binding energy per
nucleon and the charge radius for one of the much studied closed-shell nuclei, namely $^{40}$Ca. Since this nucleus is
isospin-symmetric, the model-dependence of the NM EoS plays only a minor role, if any. 
We obtain 
8.333 $\pm$ 0.200 MeV and 3.504 $\pm$ 0.077 fm 
for $B/A$ and the charge radius, respectively, to be compared with the empirical values of 
8.55 MeV and 3.48 fm. The ab initio prediction for the charge radius of
$^{40}$Ca is given in Ref.~\cite{hagen+} as 3.49(3) fm. 

In conclusion, we find that our method is able to produce realistic values for bulk nuclear properties.

\section{Conclusions and outlook}                                                                  
\label{Concl} 

The equation of state of infinite matter and its density dependence contain rich information about nucleonic
interactions in the medium, which can be extracted through the analysis of EoS-sensitive observables.
In this paper, we 
 calculated binding energies and neutron skins for neutron-rich isotopes of three selected elements 
with a method where the EoS of isospin-asymmetric matter is the crucial input. Our neutron matter EoS are based
on chiral nuclear forces constrained by $\pi N$ and NN data.                                          
In order to highlight the role of the pure neutron matter EoS, 
the calculations employed microscopic equations of state for neutron matter obtained at different orders 
of chiral EFT, whereas a phenomenological model was adopted for the EoS of symmetric 
nuclear matter.

We discussed various sources of uncertainty, paying particular attention to truncation errors. 
Predictions for both the binding energy and the neutron skin show a large truncation error at 
LO and a much smaller one at N$^3$LO. Thinking specifically of the Hamiltonian, this 
behavior is encouraging, but
complete calculations including consistent 2NF and 3NF, as well as 
calculations at N$^4$LO, will be crucial to assess a successful path to convergence.             
We also observed that the uncertainty on the energy related to a free parameter in the functional is  
typically larger than the smallest truncation error. This is not the case for the neutron skin, where 
the compounded error remains dominated by the order-by-order pattern.      
We note, further, that the 
uncertainty associated 
with this parameter is uncorrelated with the chiral order, and
so it does not hinder our ability to observe
a pattern by order, and, hopefully in the near future, a convergence pattern with respect to the Hamiltonian. 

We close by reiterating the main motivation for studies such as this one.
Our empirical knowledge of nuclear structure at the limits of stability is very limited, a
 status of affairs which is likely
to improve in the near future thanks to the development of radioactive beam facilities. 
Along with this on-going experimental efforts, it is important to carry out 
calculations based on microscopic state-of-the-art nuclear forces. 
The effective field theory approach is unique in that it allows to 
estimate the uncertainty of the predictions.                    

\section*{Acknowledgments}
This work was supported by 
the U.S. Department of Energy, Office of Science, Office of Basic Energy Sciences, under Award Number DE-FG02-03ER41270. 

\end{document}